\documentclass[11pt]{article}
\usepackage[cp1250]{inputenc}
\usepackage{amsmath}
\usepackage{amsfonts}
\usepackage{graphicx}
\usepackage{yfonts}
\usepackage{color}
\usepackage[normalem]{ulem}
\usepackage{amsmath}
\usepackage{bm}
\newcommand{\CPn}{{\mathbb{C}P^n}}
\newcommand{\CPone}{{\mathbb{C}P^1}}

\DeclareMathOperator{\imag}{i}

\DeclareMathOperator{\der}{d}

\def\<{\langle}
\def\>{\rangle}
\newtheorem{thm}{Theorem}
\newtheorem{cor}{Corollary}

\title{\bf Decoherence of a qubit as a diffusion \\ on the Bloch sphere}
\author{ Katarzyna Siudzi{\'n}ska\footnote{e-mail: kasias@stud.umk.pl } $\ $  and Dariusz Chru{\'s}ci{\'n}ski \\ Institute of Physics, Nicolaus Copernicus University, \\ ul. Grudzi\c{a}dzka 5, 87-100 Toru{\'n}, Poland }

\begin{document}

\maketitle

\begin{abstract}
We analyze qubit decoherence in the framework of geometric quantum mechanics. In this framework the qubit density operators are represented by probability distributions which are also the K\"ahler functions on the Bloch sphere. Interestingly, the complete positivity of the quantum evolution is recovered as ellipticity of the second order differential operator (deformed Laplacian) which governs the evolution of the probability distribution.
\end{abstract}


\section{Introduction}

In the geometric approach to quantum mechanics \cite{Kibble,Brody, Anandan, Ashtekar} quantum states are represented not by vectors in the Hilbert space $\mathcal{H}$, but as points in the projective Hilbert space $\mathbb{P}\mathcal{H}$. A point in $\mathbb{P}\mathcal{H}$ corresponds to a ray in $\mathcal{H}$ passing through $\psi$ or, equivalently, to the rank-1 projector $|\psi\>\<\psi|$ (see also \cite{Jamiolkowski,Bengtsson,MARMO} and \cite{Heydari} for recent reviews). If $\mathcal{H} = \mathbb{C}^{n+1}$, then
\begin{equation}\label{}
\mathbb{P}\mathcal{H} = \mathbb{C}P^{n}  = U(n+1)/U(n) \ .
\end{equation}
It turns out that $\CPn$ defines an $n$-dimensional complex space equipped with rich geometrical structures: the Fubini-Study metric $g$ and the symplectic form $\omega$, such that
\begin{equation}\label{k}
  \mathcal{K} = g + \imag\omega .
\end{equation}
The triple $(\CPn,g,\omega)$ defines a K\"ahler space (see the recent paper \cite{Heydari-JPA}). Now, for any operator $A \in \mathcal{B}(\mathcal{H})$ one defines a function $f_A : \CPn \to \mathbb{C}$ via
\begin{equation}\label{fF}
  f_A([\psi]) = \frac{\<\psi|A|\psi\>}{\<\psi|\psi\>} .
\end{equation}
These functions are called the K\"ahler functions \cite{GNS} and form a linear subspace is the space of all functions $\mathcal{F}(\CPn) :=\{ f : \CPn \to \mathbb{C} \}$. An equivalent definition of the K\"ahler function uses the following property: for any function $f : \CPn \to \mathbb{C}$  one defines the corresponding Hamiltonian vector field $X_f$ by $\der f = \omega(X_f,\, \cdot\, )$.  Now, $f$ is K\"ahlerian iff  $X_f$ is a Killing vector field, that is, $\mathfrak{L}_{X_f} g =0$. Note that if $A,B \in \mathcal{B}(\mathcal{H})$ and $C=-2\imag[A,B]$, then for the corresponding K\"ahler functions $f_A,f_B,f_C$ one has
\begin{equation}\label{}
  f_C = \{f_A,f_B\} = -X_{f_A}(f_B) \ ,
\end{equation}
where $\{f_A,f_B\} := \omega(X_{f_A},X_{f_B})$ denotes the Poisson bracket on $\CPn$ given by the symplectic form $\omega$. The corresponding Hamiltonian vector fields satisfy $X_{f_C} = [X_{f_A},X_{f_B}]$. Finally, the adjoint map ${\rm ad}_H : \mathcal{B}(\mathcal{H}) \to \mathcal{B}(\mathcal{H})$,
\begin{equation}\label{}
  {\rm ad}_H(A) = -\imag[H,A] \ ,
\end{equation}
gives rise to the map $\mathcal{F}(\CPn) \mapsto \mathcal{F}(\CPn)$ defined by $a \to \frac 12 \{h,a\}= -\frac 12 X_h(a)$.

For the qubit case the corresponding complex projective space $\mathbb{C}P^1$ is nothing but the Bloch sphere. Introducing $(\psi_0,\psi_1) \in \mathbb{C}^2$ by $$\psi_0= \cos \frac\theta 2 \ , \ \ \ \psi_1 = \sin\frac\theta 2 e^{\imag\phi} , $$
with $\theta \in [0,\pi)$ and $\phi \in [0,2\pi)$ one finds the Fubini-Study metric
\begin{equation}\label{g-2}
  \der s^2 = \frac 14 (\der\theta^2 + \sin^2\theta \der\phi^2) \ ,
\end{equation}
and the symplectic 2-form
\begin{equation}\label{omega}
  \omega = \frac 14 \sin\theta \der\theta \wedge \der\phi \ .
\end{equation}
It defines a volume element on $\CPone$ (c.f. \cite{Bengtsson})
\begin{equation}\label{}
  {\rm Vol}(\CPone) = \int_{\CPone} \omega = \pi \ .
\end{equation}
Taking the basis in $M_2(\mathbb{C})$ of $\{\sigma_0=\mathbb{I},\sigma_1,\sigma_2,\sigma_3\}$, one finds for $f_k([\psi]) = \<\psi|\sigma_k|\psi\>$:
\begin{equation}\label{}
f_0(\theta,\phi) = 1\ , \ \   f_1(\theta,\phi) = \sin\theta\cos\phi\ , \ \  f_2(\theta,\phi) = \sin\theta\sin\phi\ , \ \  f_3(\theta,\phi) = \cos\theta\ .
\end{equation}
These functions provide a basis in the space of K\"ahler functions on $\CPone$. The corresponding vector fields are ${X}_0 = 0$ and
\begin{align}\label{angular_momentum}
X_1&=-4\sin\phi\partial_\theta-4 \cot\theta\cos\phi\partial_\phi, \nonumber\\
X_2&=4\cos\phi\partial_\theta-4 \cot\theta\sin\phi\partial_\phi,\\
X_3&=4\partial_\phi\ , \nonumber
\end{align}
Note that $L_k =-\frac\imag 4 X_k$ reproduce three components of angular momentum, and hence
\begin{equation}\label{}
  L_1^2 + L_2^2 + L_3^3 = -\frac 14\Delta \ ,
\end{equation}
where $\Delta$ is the Laplacian on $\CPone$. The K\"ahler functions $\{f_1,f_2,f_3\}$ are dipole spherical harmonics, therefore $\Delta f_k = -8 f_k$.\footnote{Dipole functions $Y_{1m}$  ($m=-1,0,1$) for the Laplacian on the unit sphere satisfy $\Delta Y_{1m} = - 2 Y_{1m}$. Here, due to the factor `$\frac 14$' in (\ref{g-2}) one has `8' instead of `2'.}

In this paper we consider the above geometric structures to provide a description of decoherence processes on $\CPone$. Section \ref{QUBIT} provides a simple model of qubit decoherence, which is then analyzed within the geometric approach in Section \ref{GEO}. Section \ref{CP} discusses the conditions for complete positivity, and the final conclusions are collected in the last section.

\section{Qubit decoherence}  \label{QUBIT}

The Markovian master equation has the following form \cite{GKS,L},
\begin{equation}\label{}
  \dot{\rho}_t = - \imag[H,\rho_t] + \frac 12 \sum_k \gamma_k \left( V_k \rho_t V_k^\dagger - \frac 12 [ V_k^\dagger V_k,\rho_t]_+ \right) \ ,
\end{equation}
where $H$ denotes an effective Hamiltonian of the system, $V_k$ are the noise operators, and the coefficients $\gamma_k \geq 0$ ($[A,B]_+ = AB + BA$). Now, consider the following master equation for the evolution of a qubit,
\begin{equation}\label{ME}
  \dot{\rho}_t = \frac 12 \sum_{k=1}^3 \gamma_k ( \sigma_k \rho_t \sigma_k - \rho_t)  \ ,
\end{equation}
where $\gamma_k \geq 0$ denote the decoherence rates, and $\sigma_k$ are the Pauli matrices. The most convenient way to represent a solution of (\ref{ME}) is to use the Bloch representation,
\begin{equation}\label{}
  \rho_t = \frac 12 \left( \mathbb{I} + \sum_{k=1}^3 x_k(t) \sigma_k \right) \ .
\end{equation}
One easily finds that the Bloch vector $\mathbf{x}(t) = (x_1(t),x_2(t),x_3(t))$ satisfies
\begin{equation}\label{}
\dot{x}_1(t) = -[\gamma_2+\gamma_3]x_1(t) \ , \ \  \dot{x}_2(t) = -[\gamma_1+\gamma_3]x_2(t) \ , \ \ \dot{x}_3(t) = -[\gamma_1+\gamma_2]x_3(t) \ ,
\end{equation}
and hence
\begin{equation}\label{xxx}
  {x}_1(t) = e^{-[\gamma_2+\gamma_3]t}x_1(0) \ , \ \  {x}_2(t) = e^{-[\gamma_1+\gamma_3]t}x_2(0) \ , \ \ {x}_3(t) = e^{-[\gamma_1+\gamma_2]t}x_3(0) \ .
\end{equation}
The above solution represents an anisotropic decoherence of a qubit. Clearly, in the isotropic case, that is, if $\gamma_1=\gamma_2=\gamma_3=\gamma$, one has
$\mathbf{x}(t) = e^{-2\gamma t} \mathbf{x}(0)$, which describes a depolarizing channel. Asymptotically $x_k(t) \to 0$, which means that $\rho_t \to \frac 12 \mathbb{I}$ approaches a completely decohered maximally mixed state.

The evolution $\rho \mapsto \rho_t$ may be described by the following completely positive trace-preserving map,
\begin{equation}\label{}
  \Lambda_t[\rho]= \sum_{\alpha=0}^3 p_\alpha(t) \sigma_\alpha \rho \sigma_\alpha\
\end{equation}
with the probability distributions $p_\alpha(t)$ defined by \cite{Filip1}:
\begin{eqnarray*}
  p_1(t) &=& \frac 14 \left( 1 -  e^{-[\gamma_1+\gamma_2]t} + e^{-[\gamma_2+\gamma_3]t} - e^{-[\gamma_3+\gamma_1]t} \right)\ , \\
  p_2(t) &=& \frac 14 \left( 1 -  e^{-[\gamma_1+\gamma_2]t} - e^{-[\gamma_2+\gamma_3]t} + e^{-[\gamma_3+\gamma_1]t} \right)\ , \\
  p_3(t) &=& \frac 14 \left( 1 +  e^{-[\gamma_1+\gamma_2]t} - e^{-[\gamma_2+\gamma_3]t} - e^{-[\gamma_3+\gamma_1]t} \right) \ ,
\end{eqnarray*}
and $p_0(t) = 1 -p_1(t) -p_2(t) - p_3(t)$. Note that the Bloch vector $\mathbf{x}(t)$ stays within the Bloch ball whenever
\begin{equation}\label{ggg}
  \gamma_1 + \gamma_2 \geq 0 \ , \ \   \gamma_2 + \gamma_3 \geq 0 \ , \ \   \gamma_3 + \gamma_1 \geq 0 \ ,
\end{equation}
which is a much weaker requirement than $\gamma_1,\gamma_2,\gamma_3 \geq 0$ responsible for complete positivity of quantum evolution. Positive decoherence rates correspond to the Markovian evolution, whereas the decoherence rates satisfying (\ref{ggg}) describe the non-Markovian evolution \cite{Filip1}. We shall discuss it in more details in Section \ref{CP}.

Observe that (\ref{ME}) may be equivalently rewritten as follows,
\begin{equation}\label{ME1}
  \dot{\rho}_t = -\frac 14 \sum_{k=1}^3 \gamma_k [\sigma_k,[\sigma_k, \rho_t]]   \ .
\end{equation}
Due to the ``double commutator" structure this equation is perfectly suited for the geometric reformulation and it provides the starting point of the geometric analysis.

\section{Geometric description}  \label{GEO}

For a given density operator $\rho$ let us define
\begin{equation}\label{}
  p([\psi]) := \frac 2\pi \frac{\<\psi|\rho|\psi\>}{\<\psi|\psi\>} \ .
\end{equation}
It is clear that $p([\psi])$ is a probability distribution on $\CPone$, that is, $p([\psi]) \geq 0$ and
\begin{equation}\label{}
  \int_{\CPone} p([\psi]) \omega = 1 \ ,
\end{equation}
where $\omega$ is defined in (\ref{omega}). It corresponds to a legitimate density operators if and only if $p$ is a K\"ahler function. Note that $p$ describes a pure state $\rho=|\psi\>\<\psi|$ iff $p([\psi])=\frac 2\pi$.

For any unitary operator $U : \mathcal{H} \to \mathcal{H}$ one defines a mapping $\mathcal{U} : \mathcal{F}(\CPn) \to \mathcal{F}(\CPn)$ as follows,
\begin{equation}\label{}
  (\mathcal{U}f)([\psi]) = f([U^{-1}\psi]) \ .
\end{equation}
If $\rho_t$ satisfies the von Neumann equation
\begin{equation}\label{}
  \dot{\rho}_t = - \imag[H,\rho_t] \ ,
\end{equation}
then the corresponding family of probability distributions $p_t([\psi])$ satisfies
\begin{equation}\label{}
  \dot{p}_t=\frac 12\{h,p_t\}=-X_h\ p_t\ ,
\end{equation}
and the solution is given by
\begin{equation}\label{}
  p_t([\psi]) = p_0([e^{iHt}\psi]) \ .
\end{equation}
This solution corresponds to the rigid rotation on $\CPone$. For $H = n_0 \mathbb{I} + \sum_k n_k \sigma_k$ the initial probability distribution $p_0$ is rotated along $\mathbf{n}$ by an angle $|\mathbf{n}|t$. Now, consider the dissipative evolution of $\rho_t$ given by (\ref{ME1}).
It induces the following dynamical equation for $p_t$,
\begin{equation}\label{XXX}
  \dot{p}_t=\frac {1}{16}\sum_{k=1}^3\gamma_kX_k^2p_t \ .
\end{equation}
Interestingly, in the isotropic case $\gamma_1=\gamma_2=\gamma_3=\gamma$ one finds
\begin{equation}\label{}
  \sum_{k=1}^3 \gamma_k {X}_k^2 = 4\gamma \Delta\ ,
\end{equation}
where $\Delta$ denotes the Laplacian on $\CPone$, and hence (\ref{XXX}) reduces to the diffusion equation on the Bloch sphere,
\begin{equation}\label{D}
  \dot{p}_t=\frac{\gamma}{4}\Delta p_t \ .
\end{equation}
The diffusion equation on a sphere (\ref{D}) may be easily solved by expansion of the probability distributions into harmonic functions $Y_{lm}$. Assuming that
\begin{equation}\label{E}
  p_0 = \sum_{l=0}^\infty \sum_{m=-l}^l a_{lm} Y_{lm} \
\end{equation}
and taking into account
\begin{equation}\label{}
  \Delta Y_{lm} = - 4l(l+1) Y_{lm} \ ,
\end{equation}
one finds the solution in the following form,
\begin{equation}\label{}
  p_t = \sum_{l=0}^\infty \sum_{m=-l}^l a_{lm} e^{-\gamma l(l+1) t} Y_{lm} \ .
\end{equation}
Asymptotically $p_t \to a_{00} = \frac {1}{\pi}$. Now, if $p_t$ represents a legitimate quantum state, then the monopole and dipole functions are the only functions to enter the expansion (\ref{E}), and therefore
\begin{equation}\label{pt}
  p_t = \frac{1}{\pi}\left(1 +  e^{-2 \gamma  t} [ x_1 f_1 + x_2 f_2 + x_3 f_3] \right)   \ ,
\end{equation}
where $\mathbf{x}=(x_1,x_2,x_3)$ denotes the  Bloch vector with the initial state $\rho_0$. Note that the Bloch vector $\mathbf{x}$ corresponding to the pure state $\rho_0 = |\psi_0\>\<\psi_0|$ satisfies $|\mathbf{x}|=1$, and hence may be represented by $\mathbf{x}=(\sin\theta_0\cos\phi_0, \sin\theta_0\sin\phi_0, \cos\theta_0)$ for some point $(\theta_0,\phi_0)$. Moreover, (\ref{pt}) implies
\begin{equation}\label{pt0}
  p_t(\theta_0,\phi_0) = \frac{1}{\pi}\left(1 +  e^{-2 \gamma  t} \right)   \ ,
\end{equation}
which shows that $p_t(\theta_0,\phi_0)$ evolves from $\frac 2\pi$ at $t=0$ to $\frac 1 \pi$ as $t\to \infty$. Clearly, $  p_t(\theta_0,\phi_0) > p_t(\theta,\phi)$
for any $(\theta,\phi) \neq (\theta_0,\phi_0)$.

In the anisotropic case finding the solution of (\ref{XXX}) is much more difficult since $Y_{lm}$ are no longer the eigenfunctions of $\sum_{k=1}^3 \gamma_k X_k^2$ for $l>1$. However, for the K\"ahler functions $p_t$ one finds
\begin{equation}\label{}
  p_t = \frac{1}{\pi}\left(1 +  [e^{- [\gamma_2 + \gamma_3]t} x_1 f_1 + e^{- [\gamma_1 + \gamma_3]t} x_2 f_2 + e^{- [\gamma_1 + \gamma_2]t} x_3 f_3] \right)   \ ,
\end{equation}
which, of course, reproduces (\ref{xxx}). In the {\em axial } case, i.e. $\gamma_1=\gamma_2=0$ and $\gamma_3=\gamma$, one has
\begin{equation}\label{}
  \dot{p}_t(\theta,\phi) = \gamma \partial^2_\phi\, p_t(\theta,\phi)\ ,
\end{equation}
which is the diffusion equation on the 1-dimensional torus parameterized by $\phi$ and corresponds to the phase damping channel.

\section{Positivity vs. complete positivity} \label{CP}

Let us recall that the positivity of $\gamma_k$ provides a necessary and sufficient condition for complete positivity of the solution $\rho \mapsto \rho_t=\Lambda_t[\rho]$ of the original master equation (\ref{ME}). However, this condition is somehow missing when one considers the evolution of $p_t$ on the Bloch sphere. Note that in order to have a legitimate probability distribution provided by the K\"ahler function $p_t$ one needs to impose weaker conditions defined in (\ref{ggg}). Actually, these conditions guarantee that $\rho \mapsto \rho_t=\Lambda_t[\rho]$ is a positive map for all $t \geq 0$ \cite{Filip1}.

It should be stressed that equation (\ref{XXX}), when supplemented by (\ref{ggg}), guarantees that $p_t$ is a probability distribution on $\CPone$ if $p_0$ is a K\"ahler function. However, if $p_0$ is not K\"ahlerian (which means that it contains the multipole expansions of harmonics higher than dipole functions), then $p_t$ needs not to be a probability distribution for all $t > 0$.

Let us introduce $\Delta_\gamma := \frac 14\sum_{k=1}^3 \gamma_k {X}_k^2$ (one may call it an anisotropically deformed Laplacian).

\begin{thm} The operator $\Delta_\gamma$ is elliptic iff $\gamma_k \geq 0$ for $k=1,2,3$.
\end{thm}
Proof: The anisotropically deformed Laplacian can be rewritten as
\begin{align}\label{laplacian}
\begin{split}
\Delta_\gamma=&4\Big(\gamma_1\sin^2\phi+\gamma_2\cos^2\phi\Big)\partial_\theta^2 +4\Big(\gamma_1\cot^2\theta\cos^2\phi+\gamma_2\cot^2\theta\sin^2\phi+\gamma_3\Big)\partial_\phi^2 \\&+4\cot\theta\sin(2\phi)\Big(\gamma_1-\gamma_2\Big)\partial_\theta\partial_\phi
+4\cot\theta\Big(\gamma_1\cos^2\phi+\gamma_2\sin^2\phi\Big)\partial_\theta \\&+\frac{\sin(2\phi)}{\sin^2\theta}\Big(3+\cos(2\theta)\Big)\Big(\gamma_2-\gamma_1\Big)\partial_\phi.
\end{split}
\end{align}
Now, $\Delta_\gamma$ is elliptic on $\CPone$ iff the following diffusion matrix
\begin{equation}  \label{DIF-M}
A(\theta,\phi)=\begin{bmatrix}
a_{11}(\theta,\phi) & a_{12}(\theta,\phi) \\
a_{12}(\theta,\phi) & a_{22}(\theta,\phi)
\end{bmatrix} \ ,
\end{equation}
with the coefficients
\begin{align}\label{coefficients}
\begin{split}
a_{11}(\theta,\phi)&=4\Big(\gamma_1\sin^2\phi+\gamma_2\cos^2\phi\Big),\\
a_{22}(\theta,\phi)&=4\Big(\gamma_1\cot^2\theta\cos^2\phi+\gamma_2\cot^2\theta\sin^2\phi+\gamma_3\Big),\\ a_{12}(\theta,\phi)&=2\cot\theta\sin(2\phi)\Big(\gamma_1-\gamma_2\Big) \ ,
\end{split}
\end{align}
is positive definite for each $\theta, \phi$. Positivity of $A(\theta,\phi)$ is equivalent to the conditions
\begin{equation}\label{}
  a_{11}(\theta,\phi) \geq 0\ , \ \ a_{22}(\theta,\phi) \geq 0 \ , \ \ a_{11}(\theta,\phi) a_{22}(\theta,\phi) \geq a_{12}^2(\theta,\phi)\ .
\end{equation}
Now, $a_{11}(\theta,\phi) \geq 0$ iff $\gamma_1 \geq 0$ and $\gamma_2\geq 0$, and the condition $a_{11}(\theta,\phi) a_{22}(\theta,\phi) \geq a_{12}^2(\theta,\phi)$ is equivalent to
\begin{align}\label{results2}
&\gamma_1\gamma_2\cot^2\theta+\gamma_1\gamma_3\sin^2\phi+\gamma_2\gamma_3\cos^2\phi \geq 0 \ ,
\end{align}
which finally implies $\gamma_3 \geq 0$.   \hfill $\Box$

\begin{cor} Equation (\ref{XXX}) provides a legitimate description of the $p_t$ corresponding to a completely positive evolution of $\rho_t$ if and only if $\Delta_\gamma$ is an elliptic operator which means that (\ref{XXX}) defines the  Fokker-Planck diffusion equation on $\CPone$.
\end{cor}


\section{Conclusions}

We provide the description of qubit decoherence within geometric approach to quantum mechanics. It is shown that Markovian master equation for density operator is replaced by the diffusion-like equation for the probability distribution on the Bloch sphere. It turns out that complete positivity of the evolution is equivalent to the requirement that the evolution of the probability distribution is governed by the legitimate Fokker-Planck equation \cite{Risken}.
In the forthcoming paper we generalize the presented results to the decoherence processes on $\CPn$.

The above analysis may be easily generalized to the time-dependent case, i.e. when $\gamma_k(t)$ do depend on time. In this case, the diffusion matrix $A(\theta,\phi;t)$ defined in (\ref{DIF-M}) is time-dependent (via the time-dependence of $\gamma_k(t)$). One may characterize the Markovianity of the evolution by analyzing the properties of $A(\theta,\phi;t)$. The non-Markovian quantum evolution attracts recently considerable attention (see \cite{rev1,rev2} for the recent reviews). It turns out that memory effects might play an important role in modern quantum technologies and quantum information. Now, the evolution is CP-divisible \cite{PRL} if and only if the matrix $A(\theta,\phi;t)$ is positive-definite for all $t \geq 0$ and it is P-divisible if and only if
$A(\theta,\phi;t)$ is positive-definite on the subspace of K\"ahler functions. Hence, in the geometric approach Markovianity of the evolution is controlled by the time-dependent diffusion matrix defining the Fokker-Planck equation.

\section*{Acknowledgements}

DC was partially supported by the National Science Center project
DEC-2011/03/B/ST2/00136. We thank anonymous referees for valuable remarks.

\end{document}